%% file: Eusipco_v2_CR.tex
\documentclass[conference]{IEEEtran}
\IEEEoverridecommandlockouts
\usepackage{cite}
\usepackage{amsmath,amssymb,amsfonts,acronym}
\usepackage{algorithmic}
\usepackage{textcomp}
\usepackage[usenames]{color}
\usepackage{arydshln}
\usepackage{bm}
\usepackage{url}
\usepackage{soul}
\usepackage{comment}

\usepackage{graphicx,color}
\usepackage{pstool}

\definecolor{mycolor}{rgb}{0,0.5,0.5}
\definecolor{orange}{rgb}{1,0.5,0}

\input{acronyms}
\input{defSPAWC}

\DeclareMathOperator*{\argmin}{argmin}
\def\BibTeX{{\rm B\kern-.05em{\sc i\kern-.025em b}\kern-.08em
    T\kern-.1667em\lower.7ex\hbox{E}\kern-.125emX}}
\begin{document}

\title{Non-Centralized Navigation for Source Localization
by Cooperative UAVs
\thanks{This work has received funding from the European Union's Horizon 2020 research and innovation programme under the Marie Sklodowska-Curie project AirSens (grant no. 793581). P. M. D. thanks the support of the NSF under Award CCF-1618999.} 
}

\author{\IEEEauthorblockN{Anna Guerra $^\dagger$ \qquad Davide Dardari $^\dagger$ \qquad Petar M. Djuri\'c $^\star$ 
}
\IEEEauthorblockA{\textit{$^{\star}$ DEI-CNIT, University of Bologna, 47521, Cesena, Italy.} \\
\textit{$^{\dagger}$ ECE, Stony Brook University, Stony Brook, NY 11794, USA.}\\
E-mail: \{anna.guerra3,  davide.dardari\}@unibo.it; petar.djuric@stonybrook.edu}}

\maketitle

\begin{abstract}
In this paper, we propose a distributed solution to the navigation of a population of unmanned aerial vehicles (UAVs)  to  best localize a static source. The network is considered heterogeneous with UAVs equipped with received signal strength (RSS) sensors from which it is possible to estimate the distance from the source and/or the direction of arrival through ad-hoc rotations. This diversity in gathering and processing RSS measurements mitigates the  loss of localization accuracy due to the adoption of low-complexity  sensors. The {UAVs}  plan their trajectories on-the-fly and in a distributed fashion. The collected data are disseminated through the network via multi-hops, therefore being subject to latency. Since not all the paths are equal in terms of information gathering rewards, the motion planning is formulated as a minimization of the uncertainty of the source position under UAV kinematic and anti-collision constraints and performed by 3D non-linear programming.
The proposed analysis takes into account non-line-of-sight (NLOS) channel conditions as well as measurement age caused by the latency constraints in communication.
\end{abstract}

\bstctlcite{IEEEexample:BSTcontrol}

\section{Introduction}

In recent years, \acp{UAV} have become more and more autonomous and small, increasing the possibility of creating swarms of small flying drones able to mimic the collaborative behavior of insects \cite{wood2008first,kumar2012opportunities}.
The main motivation underlying the swarming interest is that team strategies can boost the flexibility and robustness of current wireless sensor networks. In fact, \acp{UAV} are expected to locally sense and interact with the {environment} and collaborate with each other. Moreover, the redundancy of the information coming through the network permits to lower the possibility of deterioration from the loss or malfunctioning of a single node. 
All these characteristics are exploited to enable a large number of applications \cite{bayerlein2018trajectory,zhang2015location, guerra2018collaborative,guerra2018joint}. 

In this context, the optimization of \ac{UAV} trajectories has been the subject of numerous research studies \cite{goerzen2010survey,ragi2013uav,kassas2015receding}. Among other approaches, information-seeking optimal control (i.e., strategies driven by Shannon or Fisher information measures) has been extensively investigated for localization and tracking applications \cite{ucinski2004optimal,dogancay2012uav}. For instance, in \cite{tzoreff2017path}, the problem of an off-line (pre-mission) path design for best source location using two mobile sensors is addressed. A real-time approach is proposed in \cite{shahidian2016optimal}, where multiple \acp{UAV} acquire differential \ac{RSS} measurements for ranging-based tracking. In \cite{kassas2013motion}, the motion planning is interpreted as an adaptive sensing strategy and results show the superiority of the D-optimality approach over other solutions.
\begin{figure}[t!]
\centering
\psfrag{R}[rc][rc][0.7]{\textit{Ranging}}
\psfrag{B}[lc][lc][0.7]{\textit{Bearing}}
\psfrag{J}[c][c][0.7]{\qquad \qquad\qquad \qquad \textit{Joint Rang. \& Bear.}}
\psfrag{D}[c][c][0.7]{$d$}
\psfrag{A}[c][c][0.7]{$\phi, \theta$}
\psfrag{E}[c][c][0.7]{$P_{\mathsf{rx}}$}
\psfrag{P}[c][c][0.7]{$\mathbf{p}_0$}
\psfrag{Z}[lc][lc][0.7]{RSS meas.}
 \psfrag{M}[rc][rc][0.7]{Measurement}
 \psfrag{T}[c][c][0.7]{Source}
 \psfrag{L}[rc][rc][0.7]{Link}
  \psfrag{x}[c][c][0.7]{$x$}
  \psfrag{y}[c][c][0.7]{$y$}
   \psfrag{z}[c][c][0.7]{$z$}
    \psfrag{s}[lc][lc][0.7]{$\mathbf{p}_0$}
     \psfrag{d}[rc][rc][0.7]{$d_i^{(k)}$}
     \psfrag{p}[lc][lc][0.7]{$\phi_i^{(k)}$}
     \psfrag{t}[lc][lc][0.7]{$\theta_i^{(k)}$}
     \psfrag{u}[rc][rc][0.7]{$\mathbf{p}_i^{(k)}$}
  \psfrag{C}[c][c][0.7]{\qquad\qquad Communication Link}
 \includegraphics[width=0.45\textwidth]{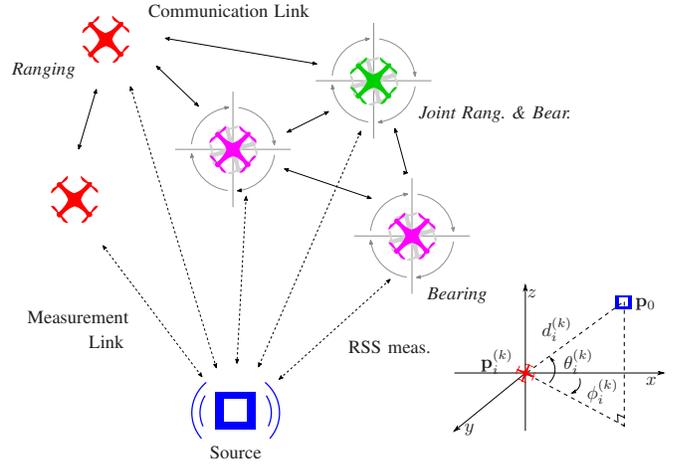}
 \caption{A UAV network, where different groups of  \acp{UAV}  acquire \ac{RSS} measurements. On the left, starting from the \ac{RSS} information, the red \acp{UAV} estimate ranging-only, the magenta bearing-only, and the green both ranging and bearing parameters. From these estimates, they navigate to best localize the source at $\mathbf{p}_0$. On the right, the coordinate system is depicted.} 
\label{fig:network}
\end{figure}

In this paper, we adopt the concept of information-seeking control and we propose a distributed navigation scheme for a network of \acp{UAV} with different sensing and processing roles, as represented in Fig.~\ref{fig:network}. Differently from the state-of-the art, a group of \acp{UAV} might infer ranging-only information while another bearing-only or both, and the obtained information is disseminated through the network via multi-hops, and hence is subjected to latency. In particular, by assuming that the UAVs know their positions thanks to on-board GPS modules, their goal is to navigate an outdoor environment for best localizing a source. 
For this purpose, the \acp{UAV} rely on measurements coming from on-board \ac{RSS} sensors, thus avoiding the use of antenna arrays, whose size and cost could be incompatible with the integration in \acp{UAV}. 

The UAV-source distance can be inferred from the  \ac{RSS} measurements and using a path-loss model \cite{dardari2012satellite}. On the other hand, the \ac{DOA} is associated with the angle from which the maximum \ac{RSS} is experienced. More specifically, \ac{UAV} rotations might be exploited to point the sensor antenna in different angular directions and to form a \ac{RSS} pattern after each rotation as in \cite{isaacs2014quadrotor}. 
Nevertheless, the time needed to search for the bearing direction and perform a complete rotation prevents drones from taking navigation decisions quickly. Hence, we suppose that not all the \acp{UAV} collect bearing measurements, opting instead for a network with heterogeneous drones. The fact that not all UAVs of the network have the possibility to estimate both ranging and bearing data might decrease the localization performance, but it also helps in reducing the time-to-navigation by avoiding UAV rotations. 

Finally, since the paths are different from an information gathering point-of-view, the problem is formulated as a 3D optimization where the function to be minimized depends on the \ac{FIM}, constrained by the \ac{UAV} kinematics and anti-collision requirements.  


\section{Problem Statement}
\label{sec:problemstate}

\subsection{UAV Dynamic Model}
We consider a network of $N$ \acp{UAV} acting as mobile reference nodes (i.e., with known positions, for instance from GPS) whose objective is to navigate through an environment to optimize the accuracy
in localizing a source in $\post=\left[\xt, \yt, \zt \right]^\text{T}$. 

At each time instant, the next position of the $i$th \ac{UAV} is given by $\fposi=\varphi\left(\posi, \mathbf{u}_i^{(k+1)} \right)$,
where $\varphi\left( \cdot \right)$ is the transition function,  $\posi=\left[\xci, \yci, \zci \right]^{\text{T}}$ is the position of the $i$th UAV at time instant $k$ and $\mathbf{u}_i^{(k)}=\left[u_{\mathsf{x},i}^{(k)}, u_{\mathsf{y},i}^{(k)}, u_{\mathsf{z},i}^{(k)}  \right]^{\text{T}}=g\left( v_i^{(k)}, \Psi_i^{(k)}, \Theta_i^{(k)}\right)$ is the control signal computed by the  $i$th \ac{UAV} on its own that enables as accurate localization of the source as possible \cite{ragi2013uav}. The speed, the heading and the tilt angles are indicated with  $v_i^{(k)}$, $\Psi_i^{(k)}$, and $\Theta_i^{(k)}$, respectively. In particular, the update for the position is given by
\begin{align}\label{eq:UAVkinematic}
    &\left[\begin{array}{l}
         x_i^{(k+1)}   \\
         y_i^{(k+1)}  \\
         z_i^{(k+1)}
    \end{array}\right] = \left[\begin{array}{l}
         x_i^{(k)} + u_{\mathsf{x},i}^{(k+1)}   \\
         y_i^{(k)} + u_{\mathsf{y},i}^{(k+1)}  \\
         z_i^{(k)} + u_{\mathsf{z},i}^{(k+1)}  
    \end{array}\right] =  \\
    &=\left[\begin{array}{l}
         x_i^{(k)}+ \left(v_{i}^{(k+1)}\cdot \Delta t \right) \, \cos\left( \Psi_i^{(k+1)} \right) \, \cos\left( \Theta_i^{(k+1)} \right) \\  
         y_i^{(k)} + \left(v_{i}^{(k+1)} \cdot \Delta t \right) \, \sin\left( \Psi_i^{(k+1)}\right) \cos\left( \Theta_i^{(k+1)} \right)   \\
          z_i^{(k)} + \left(v_{i}^{(k+1)} \cdot \Delta t \right) \, \sin\left( \Theta_i^{(k+1)}\right)  \nonumber
    \end{array}\right]\,.
\end{align}
To make the model more realistic, three constraints are added to impose the minimum and maximum speed and a maximum turn rate in both azimuthal and elevation planes \cite{dogancay2012uav}, i.e.,
\begin{equation}
\begin{cases}
         v_{\text{min}} \le \lVert \mathbf{u}_i^{(k+1)} \rVert / \Delta t \le v_{\text{max}},\\
         \lvert \Psi_i^{(k+1)}- \Psi_i^{(k)} \rvert  \leq \phi_{\text{max}}, \\
         \lvert \Theta_i^{(k+1)}- \Theta_i^{(k)} \rvert  \leq \theta_{\text{max}}, \\
\end{cases}
\end{equation}
with $\Delta t$ being the time step, $v_{\text{min}}$ and $ v_{\text{max}}$ the minimum and maximum UAV speeds, and $\phi_{\text{max}}$ and $\theta_{\text{max}}$ the turn rate limits.
The geometry of the system is depicted in Fig.~\ref{fig:network}-right.

\subsection{Observation Model}

The  \acp{UAV} obtain \ac{RSS} measurements, and from the acquired data, they extract ranging and/or bearing information from which the position of the source is estimated (two-step localization). 
Since each \ac{UAV} can process its 
measurements in a different way according to its capabilities, we indicate with $\Nranging$ the set of \acp{UAV} that obtains ranging-only estimates, with $\Nbearing$ the set that finds bearing-only estimates, and with $\Njoint$ the set with both types of estimates. The network composed of all the heterogeneous \acp{UAV} is denoted by $\Nall = \Nranging \cup \Nbearing \cup \Njoint$. 

The ranging/bearing estimation errors and the positions of the \acp{UAV} are shared through the network via multi-hops. Each node can directly communicate with its neighbors within a radius of length $\rmax$, while for greater distances, the information is delayed by $h_{ij}^{(k)}$ time slots, equal to the number of hops between the $i$th and $j$th \ac{UAV} at instant $k$.
Moreover, we assume that the connectivity is always guaranteed; as soon as a new measurement becomes available, it is recorded in an internal memory buffer and, thus, if, by chance, the UAVs get isolated, it is possible to rely on the latest saved information.

After this exchange, the data vector collected by the $i$th \ac{UAV} at time instant $k$ is $\zikvec=\left[\ldots, \tilde{\mathbf{z}}_{j}^{(\ell_k)},\, \ldots \right]^\text{T}$,
%
%
where $\tilde{\mathbf{z}}_{j}^{(\ell_k)}$ is the estimate inferred by the $j$th \ac{UAV} and arriving at the $i$th node with a delay  of $\ell_k=k-h_{ij}^{(k)}+1$. If $h_{ij}^{(k)}$ exceeds the maximum number of allowed hops (i.e., $h_\text{max}$), the \acp{UAV} refer to the last saved information. More specifically, the generic estimate is
\begin{equation}\label{eq:obs_vec2}
\tilde{\mathbf{z}}_{j}^{(\ell_k)} \!\!=\!\! \begin{cases}
 \hat{d}_j^{(\ell_k)}=d_j^{(\ell_k)}+n_{\mathsf{r},j}^{(\ell_k)}, \quad\qquad\qquad\qquad\quad\quad\,\, j \in \Nranging, \\
\hat{\alpha}_j^{(\ell_k)}= p_j^{(\ell_k)} \left({\alpha}_j^{(\ell_k)}+n_{\mathsf{b},j}^{(\ell_k)} \right) + \bar{p}_j^{(\ell_k)} \, \omega_j^{(\ell_k)},  \,\,\, j \in \Nbearing, \\
\left[ \hat{d}_j^{(\ell_k)}, \, \hat{\alpha}_j^{(\ell_k)} \right]^\text{T}, \qquad\qquad\qquad\qquad\qquad\quad\,\,\,\,\, j \in \Njoint,
\end{cases} 
\end{equation}
\noindent with $d_j^{(\ell_k)}$, $\alpha_j^{(\ell_k)} \!\!=\!\!\left(\phi_j^{(\ell_k)}, \theta_j^{(\ell_k)} \right)$ being the actual distance, azimuth and elevation angle between the $j$th \ac{UAV} and the source, and $n_{\mathsf{r},j}^{(\ell_k)} \sim \mathcal{N} \left(0, \left(\snjlk \right)^2\right)$, $n_{\mathsf{b}, j}^{(\ell_k)} \sim \mathcal{N} \left(0, \left(\sigma_{\mathsf{b},j}^{(\ell_k)}\right)^2\right)$
representing their estimation errors, respectively. The symbol $p_j^{(\ell_k)}$ is a deterministic binary variable indicating the presence or absence of \ac{NLOS}, $\bar{p}_j^{(\ell_k)}=1-p_j^{(\ell_k)}$, and $\omega_j^{(\ell_k)}$ is an outlier term due to multipath components \cite{isaacs2014quadrotor}.
The \ac{RSS}-based ranging variance can be modeled, according to the \ac{CRLB} \cite{gezici2008survey}, as
\begin{equation}\label{eq:rangingerrorRSS}
\left( \snjlk \right)^2=\left( \frac{\ln 10}{10}\cdot \frac{\sigma_{\mathsf{sh}}}{\gamma} \right)^2\, \left(d_j^{(\ell_k)} \right)^{\gamma}= \sigma^{2}_{\mathsf{r},0}\, \left( d_j^{(\ell_k)} \right)^{\gamma},
\end{equation}
\noindent where $\snok$ is the ranging \ac{std} at the reference distance ($d_{0}=1$ m),  $\gamma$ is the path-loss exponent and $\sigma_{\mathsf{sh}}$ is the shadowing \ac{std}.\footnote{The source transmit power and the channel parameters, i.e., $\sigma_{\mathsf{sh}}$ and $\gamma$, are supposed to be known. For RSS localization in presence of unknown channel parameters, we refer the reader to \cite{gholami2013rss}.} Note that the model in \eqref{eq:rangingerrorRSS} is valid for either {LOS} and \ac{NLOS} {settings} with a proper choice of the channel parameters, i.e., of the ratio ${\sigma_{\mathsf{sh}}}/{\gamma}$. 
The bearing noise variance is considered constant, i.e., $\left(\sigma_{\mathsf{b},j}^{(\ell_k)}\right)^2=\sigma^{2}_{\mathsf{b},0}$.

Starting from the collected measurements and the knowledge of the other \ac{UAV} positions, each \ac{UAV} estimates the position of the source at time instant $k$, i.e., $\epost$ \cite{dardari2012satellite}.\footnote{In this paper, we focus on the path planning aspects and we consider the position estimates given as an input for the navigation algorithm.} 

\subsection{UAV Navigation}

 Once the source position has been estimated, the \acp{UAV} plan their trajectory based on a metric capturing the quality of the localization process. One possible solution {is based on minimizing} the \ac{RMSE} of the source position estimate. {However}, the \ac{RMSE} is strictly related to  {the adopted} estimator and requires the knowledge of the actual source position (that is the parameter to be estimated) \cite{djuric2019self}, whereas the trajectory planner should be valid for any estimator, and agnostic with respect to the actual source position. To meet the requirement of invariance over estimators, we chose the following cost functions \cite{ucinski2004optimal}:
\begin{equation}\label{eq:PEB}
\mathcal{C}\left( \mathbf{q}_i^{(k)} \right)= \begin{cases}
- \ln \det \left(\mathbf{J}\left( \epost; \mathbf{q}_i^{(k)}  \right)\right)  , \,\, \text{\textit{D-Optimality}}\\
\text{tr} \left(\mathbf{J}^{-1}\left( \epost; \mathbf{q}_i^{(k)}  \right)\right) , \quad\,\,\,\,\,\, \text{\textit{A-Optimality}}
\end{cases}
\end{equation}
where $\mathbf{q}_i^{(k)}=\left[ \ldots, \mathbf{p}_j^{(\ell_k)}, \ldots\right]^\text{T}$ contains the positions of the UAVs as known by the $i$th \ac{UAV}, $\text{tr}\left( \cdot \right)$ and $\text{det}\left( \cdot \right)$ are the trace and determinant operators, respectively, and $\mathbf{J}\left(\epost; \mathbf{q}_i^{(k)}\right)$ is the \ac{FIM} of the source location as a function of the current and previous UAV locations, and evaluated on the estimated source position.
Note that  $\mathbf{q}_i^{(k)}$ can make the cost function dependent on the previous (non-updated) locations of the drones.  

Consequently, the control law at time instant $k$ at  {the $i$th} UAV is the solution of the following minimization problem:
\begin{align}\label{eq:problem}
\begin{aligned}
  \left(\mathbf{q}_i^{(k+1)}\right)^{\star}=&\argmin_{\mathbf{q}_i^{(k+1)} \in \mathbb{R}^{2}} \,\,  \mathcal{C}\left( \mathbf{q}_i^{(k+1)}\right),  \\
 &\text{subject to}
 \quad \dij \geq \dmino,\, \dik \geq \dmint,   \\
&\qquad\qquad\,\,\,\,\,\, \mathcal{T}_i \cap \mathcal{O} =  \varnothing,  \\
&\qquad\qquad\,\,\,\,\,\, v_{\text{min}} \le \lVert  \mathbf{u}_i^{(k+1)} \rVert / \Delta t \le v_{\text{max}}, \\
&\qquad\qquad\,\,\,\,\,\,  \lvert \Psi_i^{(k+1)}- \Psi_i^{(k)} \rvert  \leq \phi_{\text{max}}, \\
&\qquad\qquad\,\,\,\,\,\,  \lvert  \Theta_i^{(k+1)}- \Theta_i^{(k)} \rvert  \leq \theta_{\text{max}}, \\ 
&\qquad\qquad\,\,\,\,\,\, z_{\text{min}} \leq z_i^{(k)} \leq z_{\text{max}},
\end{aligned}
\end{align}
for $i=1, \ldots, N$, and where $\dij$ is the inter-\ac{UAV} distance, $\dmino$ is the anti-collision safety distance among \acp{UAV}, $\dmint$ {is} the safety distance with respect to the source, {$\mathcal{T}_i$ {is the set of} feasible position points of the trajectory of the $i$th \ac{UAV}}, and $\mathcal{O}$ {is} the set of obstacles present in the environment from which the \acp{UAV} should keep a safety distance {equal to} $d_{\mathsf{O}}^{*}$. The last constraints impose a maximum value on the UAV turning rates and a bounding box for the flight altitude. 

With \eqref{eq:problem}, the drones search in a fully distributed way for the optimal \ac{UAV} formation that minimizes the Fisher information-driven function at the next time instant. 

Then, recalling the transition model \eqref{eq:UAVkinematic}, the control signal of 
the $i$th \ac{UAV} that satisfies \eqref{eq:problem} is given by $\mathbf{u}_i^{(k+1)}= \left[ \left(\mathbf{q}_i^{(k+1)} \right)^{\star} \right]_i -\posi$,
%
%
where $\left[ \cdot \right]_i$ {is an operator that}  picks the $i$th entry of the optimal formation in \eqref{eq:problem}, i.e., $\left(\mathbf{p}_i^{(k+1)}\right)^{\star}$. 

\section{Cost Function Derivation}
\label{sec:costfunction}

Here we derive the cost function in \eqref{eq:problem} for both the A- and D-optimality cases. Firstly, we recall the  \ac{FIM} definition \cite{kay1998fundamentals}
\begin{align}\label{eq:FIM}
&\mathbf{J}\left(\post; \mathbf{q}_i^{(k)}\right) \!\!=\! \mathbb{E} \left\{ \left[ \nabla_{\post} \, \Lambda(\zikvec \lvert \post) \right]\!\! \left[ \nabla_{\post} \, \Lambda(\zikvec \lvert \post) \right]^\text{T} \right\}
\end{align}
with $\Lambda(\zikvec \lvert \post) \!=\! \ln f(\zikvec \lvert \post)$ being the log-likelihood function. Assuming the independence between estimates, we can write
\begin{align}\label{eq:likelihood}
  \Lambda\left( \zik \lvert \post \right) &= \sum_{j \in \Nall}   \kj \, \ln f\left(\hat{d}_j^{(\ell_k)} \lvert \post \right) +\,  \nonumber \\
  &+  \bj\, \left( \ln f\left(\hat{\theta}_j^{(\ell_k)} \lvert \post \right) + \ln f\left(\hat{\phi}_j^{(\ell_k)} \lvert \post \right)  \right) 
\end{align}
with $\kappa_{j}=\left\{0, 1\right\}$ and  $\beta_{j}=\left\{0, 1\right\}$ being equal to $1$ if the $j$th \ac{UAV} can estimate the ranging and/or bearing information. 
Following the same steps of \cite{kay1998fundamentals}, it is possible to find
\begin{align}\label{eq:classicalFIM}
&\mathbf{J}\left( \post; \mathbf{q}_i^{(k)}  \right) = \left[\begin{array}{ccc} J_{\mathsf{xx},i}^{(k)} & J_{\mathsf{xy},i}^{(k)} & J_{\mathsf{xz},i}^{(k)} \\ J_{\mathsf{xy},i}^{(k)} & J_{\mathsf{yy},i}^{(k)} &
J_{\mathsf{yz},i}^{(k)} \\
J_{\mathsf{zx},i}^{(k)} & J_{\mathsf{zy},i}^{(k)} &
J_{\mathsf{zz},i}^{(k)} \\
\end{array}  \right]  \\
&= \sum_{j \in \Nall}  \kj \,  \Ad \, \Gjlr  
+\bj \, \pjl \, \At \,  \left(\Gjlb + \Gjlbel \right), \nonumber
\end{align}
\noindent with the subscripts $\mathsf{x}$, $\mathsf{y}$, and $\mathsf{z}$ indicating the Cartesian position coordinates and where
\begin{align}\label{eq:geometricmatrix}
&\Gjlr \left(\phi_j^{(\ell_k)}, \theta_j^{(\ell_k)} \right) \!=\! {\mathbf{a}_j^{(\ell_k)} \left(  \mathbf{a}_j^{(\ell_k)} \right)^\text{T}},  \\
&\Gjlb \left( \phi_j^{(\ell_k)}, \theta_j^{(\ell_k)} \right) \!=\! \frac{\Gjlr \left(\phi_j^{(\ell_k)} + \pi/2, 0 \right)}{\left(d_j^{(\ell_k)} \, \cos \left(\theta_j^{(\ell_k)} \right) \right)^2 }, \\
&\Gjlbel \left( \phi_j^{(\ell_k)}, \theta_j^{(\ell_k)} \right) \!=\! \frac{\Gjlr \left(\phi_j^{(\ell_k)}, - \theta_j^{(\ell_k)} - \pi/2  \right)}{ \left( d_j^{(\ell_k)}   \right)^2 },
\end{align}
\noindent with $\mathbf{a}_j^{(\ell_k)}$ being the direction vector given by
\begin{equation}
\mathbf{a}_j^{(\ell_k)}=
\left[ \begin{array}{l}
\cos\left( \phi_j^{(\ell_k)} \right) \, \cos\left( \theta_j^{(\ell_k)} \right) \\
\sin\left( \phi_j^{(\ell_k)} \right) \, \cos\left( \theta_j^{(\ell_k)} \right)\\
\sin\left( \theta_j^{(\ell_k)}\right) 
\end{array}
\right].
\end{equation}
%
%
%
Finally, the coefficients $\Ad$ and $\At$ depend on the measurement noise variances as\footnote{Thanks to the possibility to discriminate LOS/NLOS situations, we assume that the \acp{UAV} exactly know the values of the coefficients in \eqref{eq:coefficients_r}.}
%
%
%
\begin{equation}\label{eq:coefficients_r}
\begin{cases}
\Ad = \left(1/\left( \snjlk \right)^2\right)\, \left( 1+2\,  \frac{\gamma^2\, \snok^2\, \left(\djl\right)^\gamma}{4\, \left(\djl\right)^2 }\right)  \\
\At ={1}/{\svok^2}
\end{cases}.
\end{equation}
%
%
%
%

Starting from \eqref{eq:classicalFIM}, the A- and D-optimality criteria can be simply derived as in \eqref{eq:PEB} where, instead of using the actual source position that is not available, the \acp{UAV} consider their estimates, i.e.,  $\epost$.
More specifically, we have
%
%
%
\begin{align}
&\mathcal{C}\left(\mathbf{q}_i^{(k)} \right)= \nonumber \\
&\begin{cases}
- \ln\left(\jxx \, C_{\mathsf{xx},i}^{(k)} + \jxy \, C_{\mathsf{yx},i}^{(k)} + \jxz \,C_{\mathsf{zx},i}^{(k)} \right)  , \,\,\, \text{\textit{D-Optimality}}\\
\frac{C_{\mathsf{xx},i}^{(k)} + C_{\mathsf{yy},i}^{(k)}+ C_{\mathsf{zz},i}^{(k)}}{\jxx \, C_{\mathsf{xx},i}^{(k)} + \jxy \, C_{\mathsf{yx},i}^{(k)} + \jxz \,C_{\mathsf{zx},i}^{(k)} }, \qquad\qquad\quad\qquad \text{\textit{A-Optimality}} \nonumber
\end{cases}
\end{align}
with the cofactors of the \ac{FIM} given by $C_{\mathsf{xx}} \!=\! \jyyi \jzzi \!-\! \left(\jyzi \right)^2$, $C_{\mathsf{yy}} \!=\! \jxxi \jzzi \!-\! \left(\jxzi\right)^2 $, $C_{\mathsf{yx}} \!=\!  \jyzi \jxzi \!-\! \jxyi \jzzi$, $C_{\mathsf{zx}} \!=\! \jxyi \jyzi \!-\! \jyyi  \jxzi$, and $C_{\mathsf{zz}} \!=\! \jxxi \jyyi \!-\! \left( \jxyi \right)^2$.
\section{Control Law}
\label{sec:controllaw}

The constrained minimization problem in \eqref{eq:problem} can be solved using the projection gradient method  \cite{luenberger1984linear}
\begin{equation}\label{eq:update_g2}
\mathbf{u}_i^{(k+1)} \!=\! -\xi\, \mathbf{P}\, \nabla_{\mathbf{p}_{i}^{(k)}} \, \mathcal{C}\left( \mathbf{q}_i^{(k)}\right) \!-\! \mathbf{N}\left(\mathbf{N}^\text{T} \mathbf{N}\right)^{-1}\mathbf{g},
\end{equation}
where $\xi$ represents the spatial step,  {$\nabla_{\mathbf{p}_{i}^{(k)}} \, \left(\cdot \right)$} is the gradient operator with respect to the \ac{UAV} positions which, taken with the negative sign, represents the direction of decrease of the cost function. 
The projection matrix is denoted with $\mathbf{P}=\mathbf{I}-\mathbf{N}\left(\mathbf{N}^\text{T} \mathbf{N}\right)^{-1}\mathbf{N}^\text{T}$ with $\mathbf{I}$ being the identity matrix and $\mathbf{N}=\left( \nabla_{\mathbf{p}_{i}^{(k)}}\mathbf{g} \right)$ being the gradient of the constraints in $\mathbf{g}=\left[\mathbf{g}_{1} \,\, \mathbf{g}_{2} \,\, \mathbf{g}_{3} \right]$, where 
\begin{align}
&\mathbf{g}_{1}=\mathbf{d}_{\text{U}}-\dmino, \quad \mathbf{d}_{\text{U}}=\left\{\dij: \dij< \dmino\right\}, \\
&\mathbf{g}_{2}=\mathbf{d}_{\text{S}}-\dmint, \quad\,\, \mathbf{d}_{\text{S}}=\left\{\di^{(k)}: \di^{(k)}< \dmint\right\}, \\
&\mathbf{g}_{3}=\mathbf{d}_{\text{O}}-d_{\text{O}}^*, \quad\,\, \mathbf{d}_{\text{O}}=\left\{d_{i, \text{O}}^{(k)}: d_{i, \text{O}}^{(k)}< d_{\text{O}}^*\right\},
\end{align}
%
with $d_{i, \text{O}}^{(k)}$ being the minimum distance between the $i$th drone and its closest obstacle.
Finally, we impose the \ac{UAV} speed, altitude and the maximum turning rates by considering the last four constraints of \eqref{eq:problem}.

\section{Case Study}
\label{sec:cs}
\begin{figure}[t!]

    \centering
\includegraphics[width=0.23\textwidth]{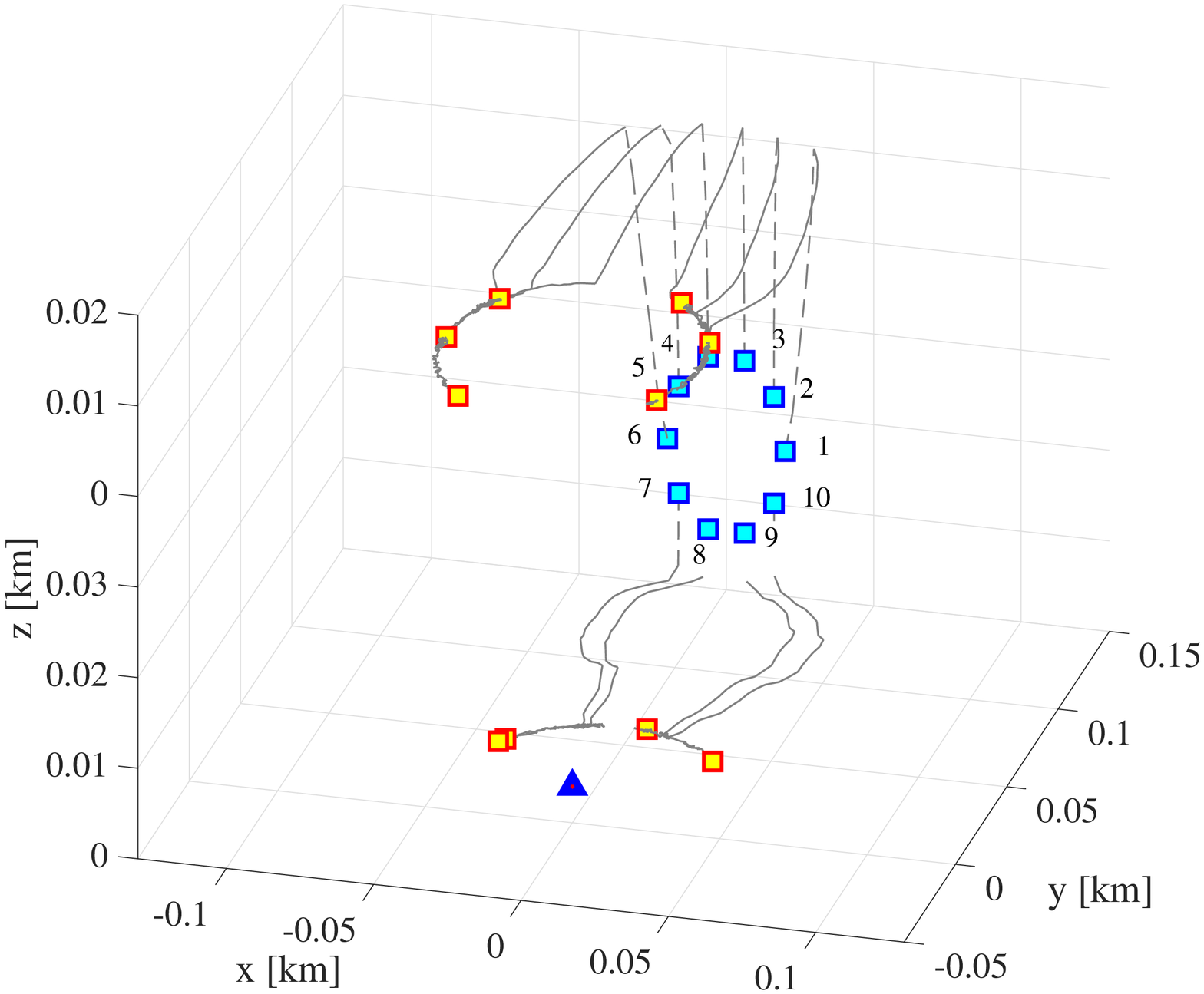}
\includegraphics[width=0.23\textwidth]{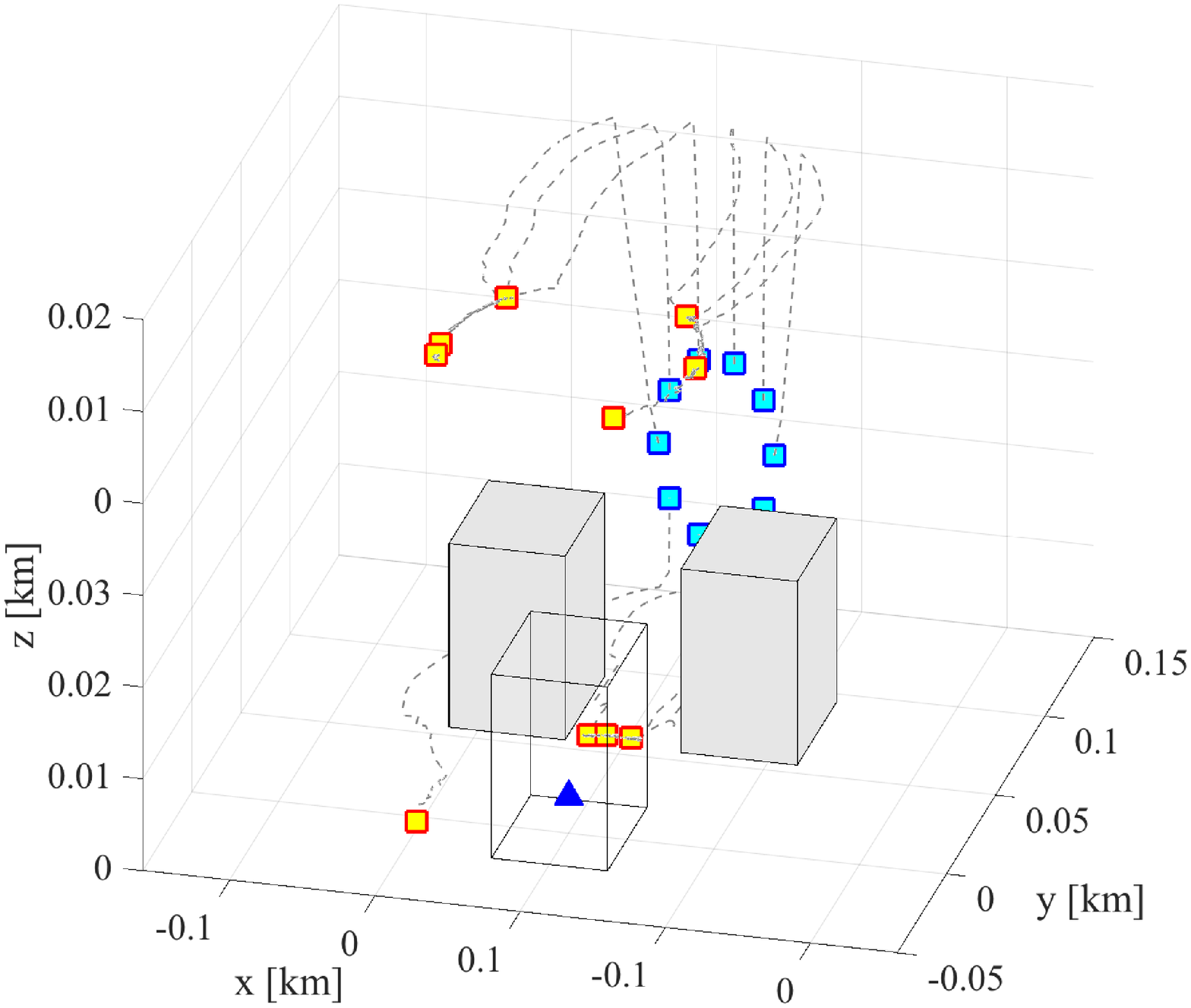} \\
\includegraphics[width=0.23\textwidth]{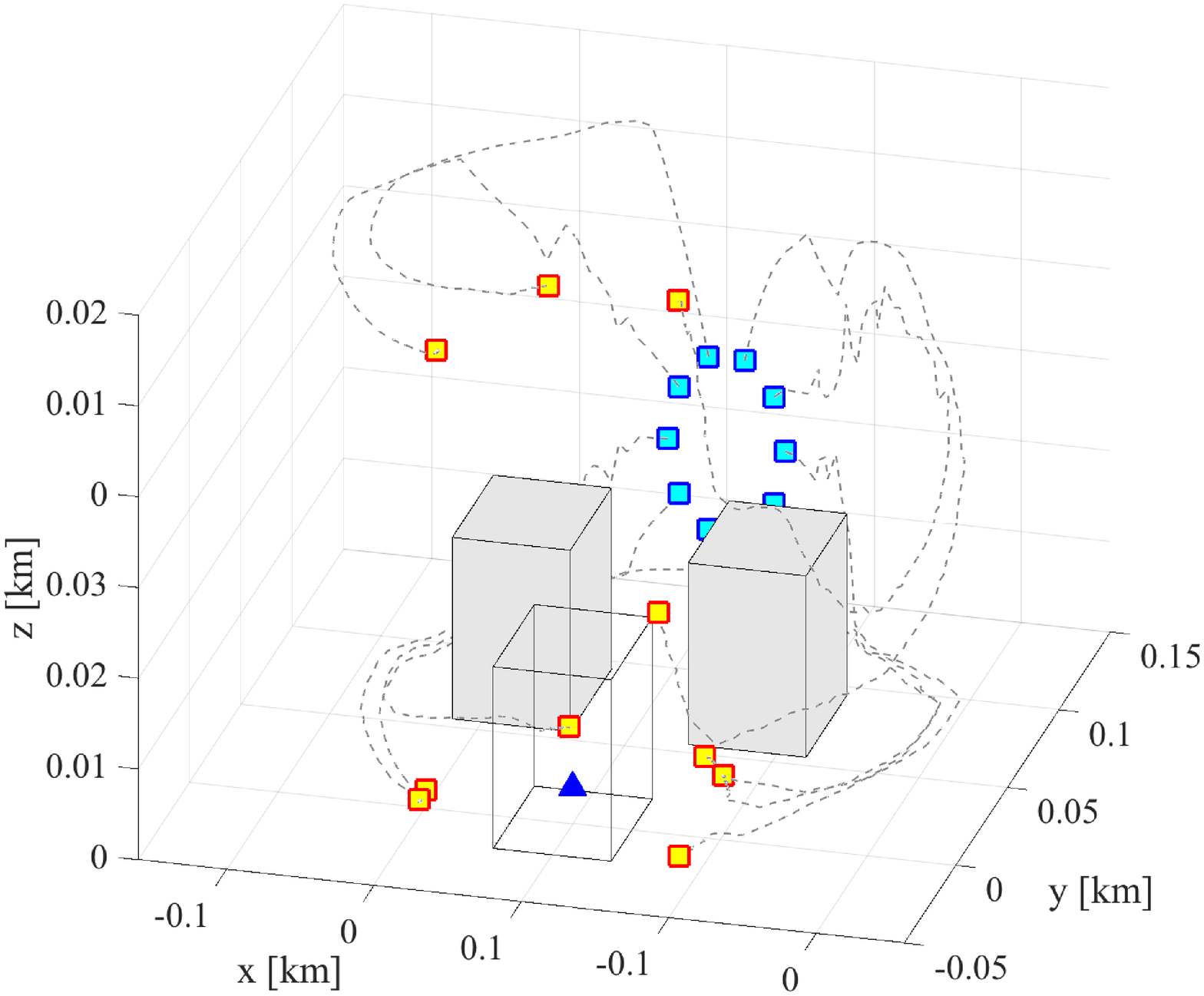}
\includegraphics[width=0.23\textwidth]{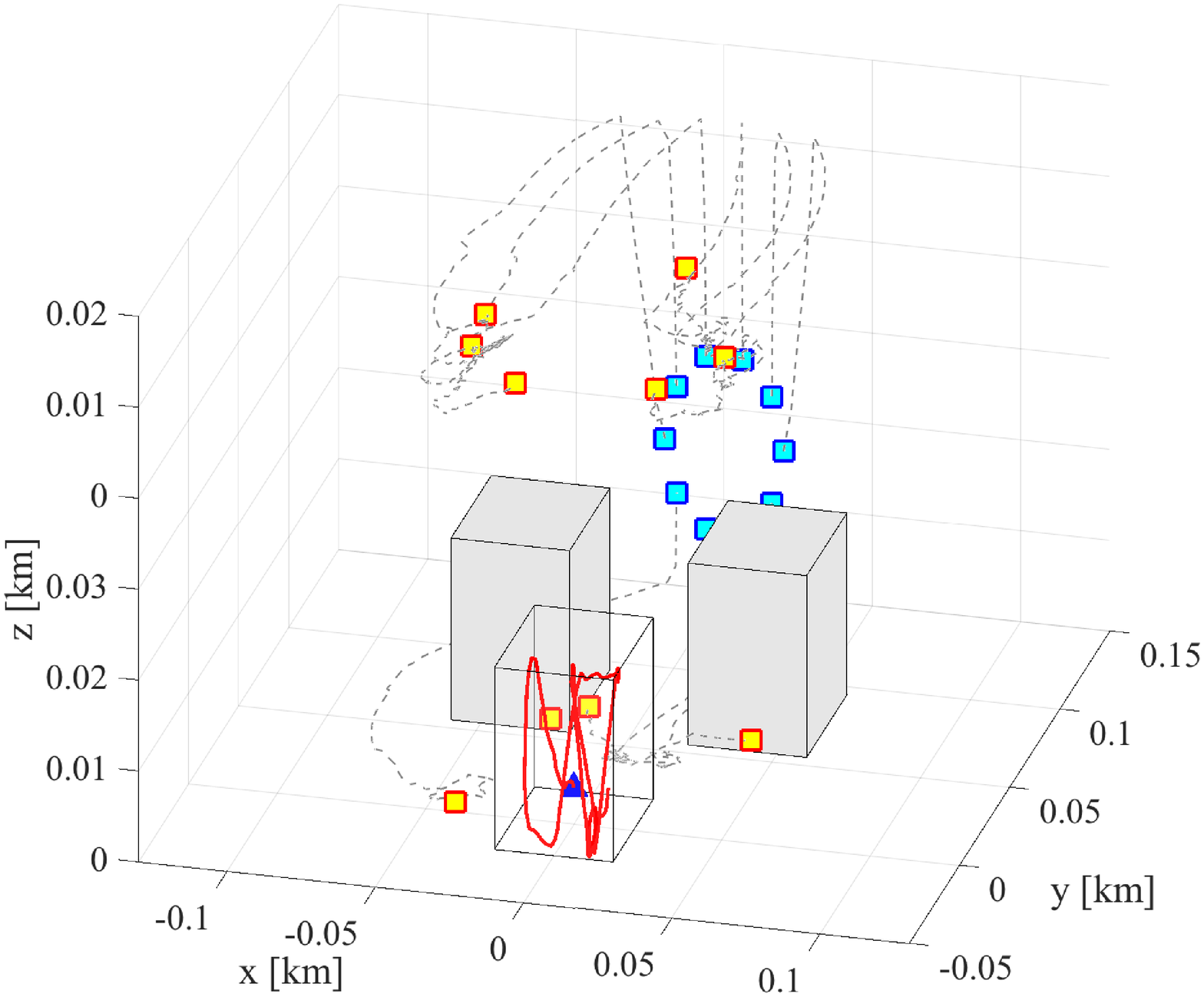}\\
\includegraphics[width=0.23\textwidth]{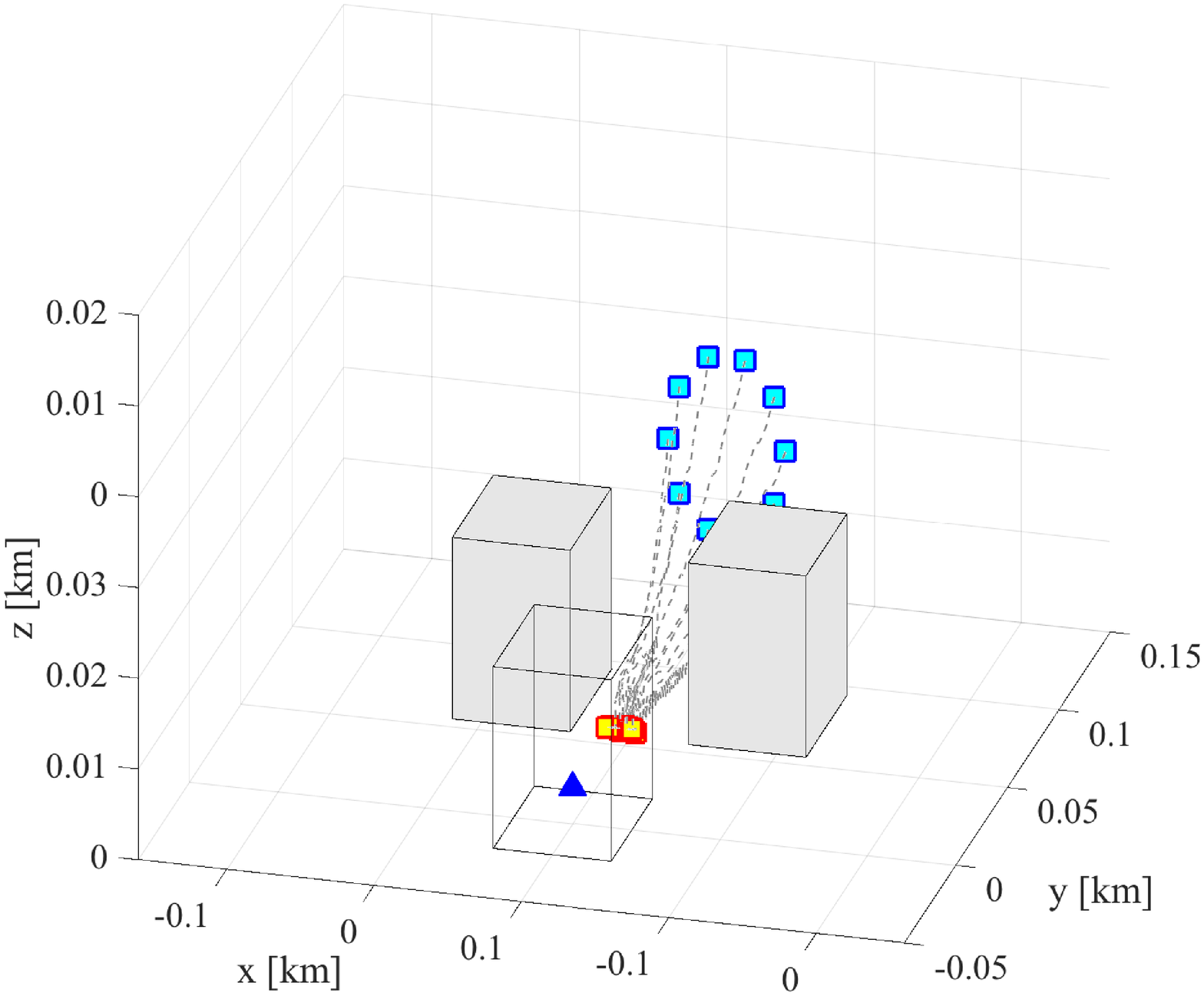}
\includegraphics[width=0.23\textwidth]{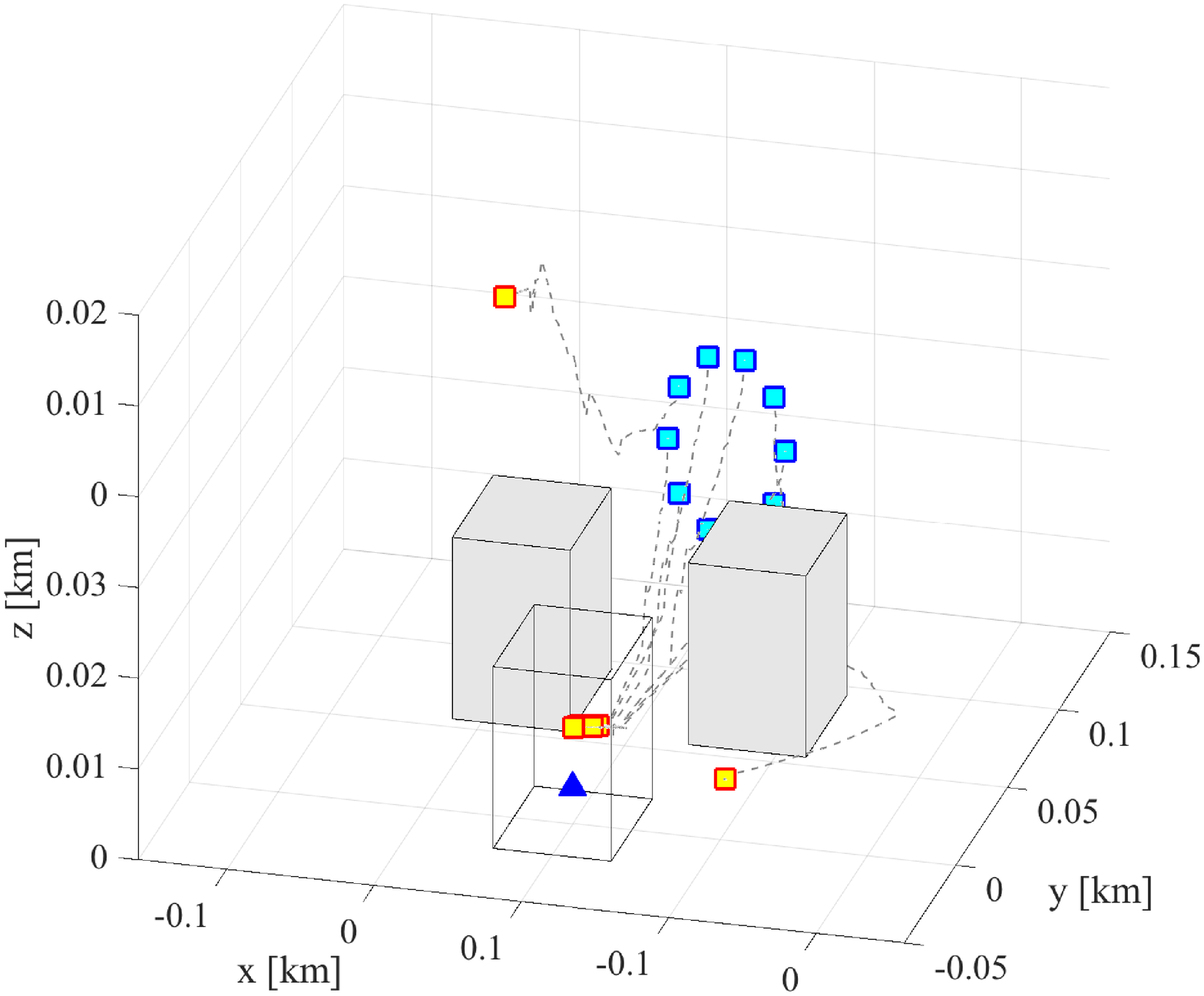}
\caption{Example of UAV trajectories obtained with the A-optimality criterion. Top-left: $\Nall=\Nranging$, LOS scenario; Top-Right: $\Nall=\Nranging$, NLOS scenario; Middle-left: $\Nall=\Nbearing$; Middle-right: $\Nall=\Nranging$, with a dynamic source; Bottom-left: $\Nall=\Njoint$; Bottom-right: $\Nall=\Nranging \cup \Nbearing \cup \Njoint$.}
\label{fig:rangingbearing_scenario}
\end{figure}

In this section, we analyze the evolution of the localization accuracy in relation to the \ac{UAV} dynamics and sensing capabilities for a 3D scenario plotted in Fig.~\ref{fig:rangingbearing_scenario}.

The minimum and maximum speed were set to $v_{\text{min}}=0.5\,$m/step and $v_{\text{max}}=1\,$m/step, respectively, while the maximum turn rates per unit step to $\phi_{\text{max}}= \theta_{\text{max}}=50^\circ$. The estimation errors were $\sigma_{\mathsf{sh}}/\gamma=1.7$ for LOS, $\sigma_{\mathsf{sh}}/\gamma=3.2$ for NLOS \cite{gezici2008survey}, and $\svok=10^\circ$ \cite{isaacs2014quadrotor}. The safety distances were $\dmino=1\,$m, $\dmint=50\,$m and $d_\text{O}^*=5\,$m. The range of altitudes of the UAVs was set to $\left[z_{\text{min}}, z_{\text{max}} \right]=\left[2, 25 \right]\,$m. {There were 100  Monte Carlo trials where at each iteration a different measurement noise was generated}, with $N=10$, a communication range of $\rmax=100\,$m, and $\hmax=1$. Each Monte Carlo simulation was restricted to $550$ steps. In the heterogeneous case, we set $\Nranging=\left\{1,4,7,10 \right\}$, $\Nbearing=\left\{2,5,8 \right\}$ and $\Njoint=\left\{3,6,9 \right\}$ with the numbering reported in Fig.~\ref{fig:rangingbearing_scenario}-top-left. The initial positions of UAVs was set along an ellipse in the $XZ-$plane of radii $r_{\mathsf{x}}= 20$ m, $r_{\mathsf{z}}= 5$m and centered at $[0, 150, 8]$ m.

In Fig.~\ref{fig:rangingbearing_scenario}, the \ac{UAV} positions are displayed as squares of different colors (for time instants $k= \left\{0, 550 \right\}$), the source position is the blue triangle. The trajectories are drawn as dashed lines, and the obstacles creating NLOS situations are the grey parallelepipeds. Figure~\ref{fig:rangingbearing_scenario} displays some qualitative examples of trajectories estimated using the  A-optimality criterion. More specifically, we considered four cases depending on the information collected by the \acp{UAV}: 
\begin{itemize}
\item The ranging-only case, when $\mathcal{N}=\Nranging$, as in Fig.~\ref{fig:rangingbearing_scenario}-top; \item The bearing-only case, when  $\mathcal{N}=\Nbearing$, as in Fig.~\ref{fig:rangingbearing_scenario}-middle-left;
\item The joint  ranging and bearing case,  when $\mathcal{N}=\Njoint$, as in Fig.~\ref{fig:rangingbearing_scenario}-bottom-left;
\item The case with heterogeneous \acp{UAV}, i.e., $\mathcal{N}=\Nranging \cup \Nbearing \cup \Njoint$, as in Fig.~\ref{fig:rangingbearing_scenario}-bottom-right. 
\end{itemize}
%
In all the simulations, the source was supposed to be static. Anyway, in Fig.~\ref{fig:rangingbearing_scenario}-middle-right, an example with a dynamic source is reported. As can be noticed, due to the fact that the movements are constrained by the perimeter, there are no significant differences compared to  Fig.~\ref{fig:rangingbearing_scenario}-top-right.

\begin{figure}[t!]
    \centering
     \psfrag{PEB}[c][c][0.8]{$\overline{\mathsf{PEB}}^{\,(k)}$ [m]}
      \psfrag{Steps}[c][c][0.8]{Steps $k$}
      \psfrag{RagingOnly}[lc][lc][0.7]{$\Nall=\Nranging$}
      \psfrag{BearingOnly}[lc][lc][0.7]{$\Nall=\Nbearing$}
      \psfrag{JointBearingRanging}[lc][lc][0.7]{$\Nall=\Njoint$}
      \psfrag{Heterogeneous}[lc][lc][0.7]{$\Nall=\Nranging \cup \Nbearing \cup \Njoint$}
\includegraphics[width=0.43\textwidth]{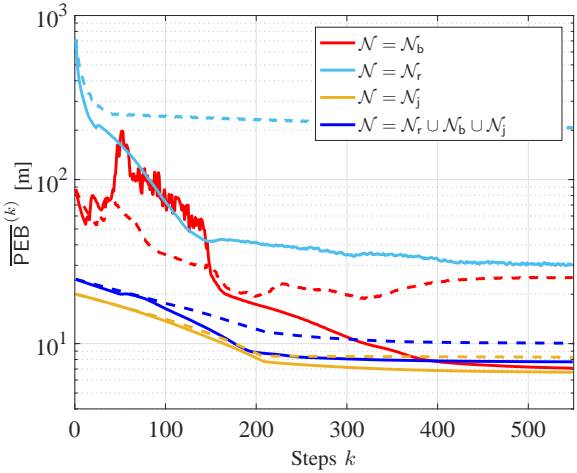}
    \caption{\ac{PEB} \textit{vs.} type of measurements with $N=10$, $\rmax=100$ m and $\hmax=1$. The dashed lines refer to the D-optimality criterion whereas the continuous lines to the A-optimality approach.}
    \label{fig:fig_typemeas}
\end{figure}

To assess the localization accuracy, we considered the \ac{PEB} averaged over the Monte Carlo trials, defined as
\begin{equation}
  \overline{\mathsf{PEB}}^{(k)}=\frac{1}{N \cdot \Nmc} \sum_{i=1}^{N} \sum_{m=1}^{\Nmc}  \sqrt{\text{tr} \left(\mathbf{J}^{-1}\left( \post; \mathbf{q}_{i,m}^{(k)}  \right)\right) },  
\end{equation}
where $\mathbf{q}_{i,m}^{(k)}$ represents the set of \acp{UAV} positions available at the $i$th \ac{UAV} at time instant $k$ at the $m$th Monte Carlo iteration.

We will show that when the swarm is composed of heterogeneous \acp{UAV}, the performance in terms of localization accuracy is similar to that obtainable with \acp{UAV} acquiring both  ranging and bearing information. 

In Fig.~\ref{fig:fig_typemeas}, the \ac{PEB} is {presented} as a function of the criterion used for the navigation (i.e., A- and D-optimality) and considering different \ac{UAV} sensing capabilities. 
As can be observed, when the \acp{UAV} {have} bearing-only measurements, {in \ac{NLOS} settings, 
obstacle obstructions lead to a singular \ac{FIM}, which might prevent the \acp{UAV} from  further navigation}. The step behaviour of the \ac{PEB} curve in the bearing-only case is due to this effect. Namely, before becoming a singular matrix, the \ac{FIM} decreases (and, hence, the \ac{PEB} increases) because some \acp{UAV} have already entered the NLOS area and their measurements are no longer related with the source position. To avoid that the \acp{UAV} stop due to a non-sufficient number of informative measurements, a ``random" approach is adopted where the \acp{UAV} move randomly along the last followed direction.
On the contrary, for the ranging-only case, the effect of the \ac{NLOS} is an increased ranging error variance. These issues are solved when the \acp{UAV} can {obtain} both ranging and bearing measurements or when the network is composed of heterogeneous \acp{UAV}. 
An interesting point emerging from Fig.~\ref{fig:fig_typemeas} is that, for navigation in \ac{NLOS} environments, the A-optimality criterion outperforms the D-optimality in terms of localization accuracy  for most of the navigation time. In fact, when using the  D-optimality criterion, the \acp{UAV} are trapped in the \ac{NLOS} areas longer than when the A-optimality is used. 
\section{Conclusions}
\label{sec:conclusions}

In this paper, we proposed a distributed network of heterogeneous \acp{UAV} whose navigation task is to minimize the localization error of a static source. In particular, the proposed control law aims at minimizing two Fisher information-driven cost functions (i.e., A- and D-optimality functions) in a distributed fashion and at each time instant. The cost functions were derived under the assumption of having access to bearing and/or ranging estimates.  The results demonstrate that multi-modal \acp{UAV} adopting the A-optimality criterion achieve the best localization performance for the considered scenario.

\bibliographystyle{IEEEtran}

\end{document}

%% file: acronyms.tex
\acrodef{BIM}{Bayesian information matrix}

\acrodef{CRLB}{Cram\'er-Rao lower bound}

\acrodef{DOA}{direction-of-arrival}
\acrodef{EKF}{Extended Kalman Filter}
\acrodef{FANET}{flying ad-hoc network} 
\acrodef{FIM}{Fisher Information Matrix}
\acrodef{FMCW}{frequency-modulated continuous-wave}
\acrodef{GPS}{Global Positioning System}
\acrodef{LOS}{line-of-sight}
\acrodef{MILP}{Mixed Integer Linear Programming}
\acrodef{MCP}{Model Predictive Control}
\acrodef{MAV}{micro aerial vehicle}
\acrodef{ML}{maximum likelihood}
\acrodef{MSE}{mean squared error}
\acrodef{NLOS}{non line-of-sight}
\acrodef{NFER} {Near-Field Electromagnetic Ranging}
\acrodef{PEB}{Position Error Bound}
\acrodef{RV}{random variable}
\acrodef{RSS}{received signal strength}
\acrodef{RMSE}{root mean squared error}
\acrodef{RCS}{radar cross section}
\acrodef{RTT}{round-trip time}
\acrodef{SLAM}{Simultaneous Localization and Mapping}
\acrodef{std}{standard deviation}
\acrodef{SNR}{signal-to-noise ratio}
\acrodef{TOA}{time of arrival}
\acrodef{UAV}{unmanned aerial vehicle}
\acrodef{UWB}{ultrawide bandwidth}
\acrodef{WSN}{wireless sensor network}
\acrodef{WRN}{wireless radar network}

%% file: defSPAWC.tex
\newcommand{\posi} {\mathbf{p}_{i}^{(k)}}

\newcommand{\fposi} {\mathbf{p}_{i}^{(k+1)}}
\newcommand{\xci} {x_{i}^{(k)}}
\newcommand{\yci} {y_{i}^{(k)}}
\newcommand{\zci} {z_{i}^{(k)}}

\newcommand{\post} {\mathbf{p}_{0}}

\newcommand{\epost} {\hat{\mathbf{p}}_{0\, i}^{(k)}}

\newcommand{\xt} {x_{0}}
\newcommand{\yt} {y_{0}}
\newcommand{\zt} {z_{0}}

\newcommand{\rmax} {d_\mathsf{hop}}

\newcommand{\zikvec} {\mathbf{z}_{i}^{(k)}}

\newcommand{\dik} {d_{i}^{(k)}}
\newcommand{\di} {d_i}
\newcommand{\dij} {d_{ij}^{(k)}}

\newcommand{\dmino} {d_\text{U}^*}
\newcommand{\dmint} {d_\text{S}^*}

\newcommand{\Nmc} {N_\text{MC}}

\newcommand{\hmax} {h_\text{max}}

\newcommand{\Nranging}{\mathcal{N}_\mathsf{r}}
\newcommand{\Nbearing}{\mathcal{N}_\mathsf{b}}
\newcommand{\Njoint}{\mathcal{N}_\mathsf{j}}
\newcommand{\Nall} {\mathcal{N}}

\newcommand{\snjlk}{\sigma_{\mathsf{r},j}^{(\ell_k)}}
\newcommand{\snok}{\sigma_{\mathsf{r},0}}

\newcommand{\svok}{\sigma_{\mathsf{b},0}}

\newcommand{\zik}{\mathbf{z}_i^{(k)}}

\newcommand{\djl}{d_j^{(\ell_k)}}

\newcommand{\pjl}{p_j^{(\ell_k)}}

\newcommand{\kj}{\kappa_j}
\newcommand{\bj}{\beta_j}

\newcommand{\Gjlr}{\mathbf{G}_{\mathsf{r},j}^{(\ell_k)}}
\newcommand{\Gjlb}{\mathbf{G}_{\mathsf{az},j}^{(\ell_k)}}
\newcommand{\Gjlbel}{\mathbf{G}_{\mathsf{el},j}^{(\ell_k)}}

\newcommand{\Ad}{A_{\mathsf{r},j}^{(\ell_k)}}
\newcommand{\At}{A_{\mathsf{b},j}^{(\ell_k)}}

\newcommand{\jxx}{J_{\mathsf{xx},i}^{(k)}}
\newcommand{\jxy}{J_{\mathsf{xy},i}^{(k)}}
\newcommand{\jxz}{J_{\mathsf{xz},i}^{(k)}}

\newcommand{\jxxi}{J_{\mathsf{xx}}}
\newcommand{\jxyi}{J_{\mathsf{xy}}}
\newcommand{\jxzi}{J_{\mathsf{xz}}}
\newcommand{\jyyi}{J_{\mathsf{yy}}}
\newcommand{\jyzi}{J_{\mathsf{yz}}}
\newcommand{\jzzi}{J_{\mathsf{zz}}}

%% file: Eusipco_v2_CR.bbl
\begin{thebibliography}{10}
\providecommand{\url}[1]{#1}
\csname url@samestyle\endcsname
\providecommand{\newblock}{\relax}
\providecommand{\bibinfo}[2]{#2}
\providecommand{\BIBentrySTDinterwordspacing}{\spaceskip=0pt\relax}
\providecommand{\BIBentryALTinterwordstretchfactor}{4}
\providecommand{\BIBentryALTinterwordspacing}{\spaceskip=\fontdimen2\font plus
\BIBentryALTinterwordstretchfactor\fontdimen3\font minus
  \fontdimen4\font\relax}
\providecommand{\BIBforeignlanguage}[2]{{%
\expandafter\ifx\csname l@#1\endcsname\relax
\typeout{** WARNING: IEEEtran.bst: No hyphenation pattern has been}%
\typeout{** loaded for the language `#1'. Using the pattern for}%
\typeout{** the default language instead.}%
\else
\language=\csname l@#1\endcsname
\fi
#2}}
\providecommand{\BIBdecl}{\relax}
\BIBdecl

\bibitem{wood2008first}
R.~J. Wood, ``The first takeoff of a biologically inspired at-scale robotic
  insect,'' \emph{IEEE Trans. Robot.}, vol.~24, no.~2, pp. 341--347, 2008.

\bibitem{kumar2012opportunities}
V.~Kumar and N.~Michael, ``Opportunities and challenges with autonomous micro
  aerial vehicles,'' \emph{The Int. J. Robot. Research}, vol.~31, no.~11, pp.
  1279--1291, 2012.

\bibitem{bayerlein2018trajectory}
H.~Bayerlein, P.~De~Kerret, and D.~Gesbert, ``Trajectory optimization for
  autonomous flying base station via reinforcement learning,'' in \emph{Proc.
  2018 19th Int. Workshop Signal Process. Adv. Wireless Commun. (SPAWC)}.\hskip
  1em plus 0.5em minus 0.4em\relax IEEE, 2018, pp. 1--5.

\bibitem{zhang2015location}
S.~Zhang \emph{et~al.}, ``Location-aware formation control in swarm
  navigation,'' in \emph{Proc. 2015 Globecom Workshops}.\hskip 1em plus 0.5em
  minus 0.4em\relax IEEE, 2015, pp. 1--6.

\bibitem{guerra2018collaborative}
A.~Guerra \emph{et~al.}, ``Collaborative target-localization and
  information-based control in networks of {UAV}s,'' in \emph{Proc. 19th Int.
  Workshop Signal Process. Adv. Wireless Commun. (SPAWC)}.\hskip 1em plus 0.5em
  minus 0.4em\relax IEEE, 2018, pp. 1--5.

\bibitem{guerra2018joint}
A.~Guerra, D.~Dardari, and P.~M. Djuri{\'c}, ``Joint indoor localization and
  navigation of uavs for network formation control,'' in \emph{Proc. 2018 52nd
  Asilomar Conf. Signals, Sys., and Comput.}\hskip 1em plus 0.5em minus
  0.4em\relax IEEE, 2018, pp. 13--19.

\bibitem{goerzen2010survey}
C.~Goerzen, Z.~Kong, and B.~Mettler, ``A survey of motion planning algorithms
  from the perspective of autonomous {UAV} guidance,'' \emph{J. Intell. Robotic
  Syst.}, vol.~57, no. 1-4, p.~65, 2010.

\bibitem{ragi2013uav}
S.~Ragi and E.~K. Chong, ``{UAV} path planning in a dynamic environment via
  partially observable {M}arkov decision process,'' \emph{IEEE Trans. Aerosp.
  Electron. Syst.}, vol.~49, no.~4, pp. 2397--2412, 2013.

\bibitem{kassas2015receding}
Z.~M. Kassas and T.~E. Humphreys, ``Receding horizon trajectory optimization in
  opportunistic navigation environments,'' \emph{IEEE Trans. Aerosp. Electron.
  Syst.}, vol.~51, no.~2, pp. 866--877, 2015.

\bibitem{ucinski2004optimal}
D.~Ucinski, \emph{Optimal measurement methods for distributed parameter system
  identification}.\hskip 1em plus 0.5em minus 0.4em\relax CRC Press, 2004.

\bibitem{dogancay2012uav}
K.~Dogancay, ``{UAV} path planning for passive emitter localization,''
  \emph{IEEE Trans. Aerosp. Electron. Syst.}, vol.~48, no.~2, pp. 1150--1166,
  2012.

\bibitem{tzoreff2017path}
E.~Tzoreff and A.~J. Weiss, ``Path design for best emitter location using two
  mobile sensors,'' \emph{IEEE Trans. Signal Process.}, vol.~65, no.~19, pp.
  5249--5261, 2017.

\bibitem{shahidian2016optimal}
S.~A.~A. Shahidian and H.~Soltanizadeh, ``Optimal trajectories for two {UAV}s
  in localization of multiple {RF} sources,'' \emph{Trans. Institute Meas.
  Control}, vol.~38, no.~8, pp. 908--916, 2016.

\bibitem{kassas2013motion}
Z.~Kassas and T.~E. Humphreys, ``Motion planning for optimal information
  gathering in opportunistic navigation systems,'' in \emph{AIAA Guidance,
  Navig., Control (GNC) Conf.}, 2013, p. 4551.

\bibitem{dardari2012satellite}
D.~Dardari, E.~Falletti, and M.~Luise, \emph{Satellite and terrestrial radio
  positioning techniques: a signal processing perspective}.\hskip 1em plus
  0.5em minus 0.4em\relax Elsevier, 2012.

\bibitem{isaacs2014quadrotor}
J.~T. Isaacs \emph{et~al.}, ``Quadrotor control for {RF} source localization
  and tracking,'' in \emph{Proc. 2014 Int. Conf. {U}nmanned {A}ircraft {S}ys.
  (ICUAS)}.\hskip 1em plus 0.5em minus 0.4em\relax IEEE, 2014, pp. 244--252.

\bibitem{gezici2008survey}
S.~Gezici, ``A survey on wireless position estimation,'' \emph{Wireless
  personal communications}, vol.~44, no.~3, pp. 263--282, 2008.

\bibitem{gholami2013rss}
M.~R. Gholami, R.~M. Vaghefi, and E.~G. Str{\"o}m, ``{RSS}-based sensor
  localization in the presence of unknown channel parameters,'' \emph{IEEE
  Trans. Signal Process.}, vol.~61, no.~15, pp. 3752--3759, 2013.

\bibitem{djuric2019self}
P.~M. Djuri\'c and P.~Closas, ``On self-assessment of proficiency of autonomous
  systems,'' in \emph{Proc. 2019 Int. Conf. Acoustics, Speech and Signal
  Process. (ICASSP)}.\hskip 1em plus 0.5em minus 0.4em\relax IEEE, 2019, pp.
  5072--5076.

\bibitem{kay1998fundamentals}
S.~M. Kay \emph{et~al.}, \emph{Fundamentals of statistical signal processing,
  Vol. {I}: Estimation theory}.\hskip 1em plus 0.5em minus 0.4em\relax Prentice
  Hall Upper Saddle River, NJ, USA:, 1998.

\bibitem{luenberger1984linear}
D.~G. Luenberger, Y.~Ye \emph{et~al.}, \emph{Linear and nonlinear
  programming}.\hskip 1em plus 0.5em minus 0.4em\relax Springer, 1984, vol.~2.

\end{thebibliography}
